\documentclass[twocolumn, aps, pre, floatfix, showpacs]{revtex4}
\usepackage{graphicx}

\newif\iffigs
\figstrue   
\iffigs
\fi

\def\drawing #1 #2 #3 {
\begin{center}
\setlength{\unitlength}{1mm}
\begin{picture}(#1,#2)(0,0)
\put(0,0){\framebox(#1,#2){#3}}
\end{picture}
\end{center} }

\newcommand{\eqref}[1]{(\ref{#1})}

\newcommand{\secref}[1]{Section~\ref{#1}}
\newcommand{\figref}[1]{Fig.~\ref{#1}}
\renewcommand{\vec}[1]{\mbox{\boldmath $#1$}}
\newcommand{\tomega}{\tilde{\omega}}
\newcommand{\chiy}{\chi_y}
\newcommand{\mPsi}{\mit\Psi}
\newcommand{\mOmega}{\mit\Omega}
\newcommand{\mTheta}{\mit\Theta}

\begin{document}
\title{A class of steady solutions to two-dimensional free convection}
\thanks{submitted to {\it Phys.Rev.E}}
\author{Sadayoshi \surname{Toh}}
\author{Takeshi \surname{Matsumoto}}
\affiliation{Division of Physics and Astronomy, Graduate School of Science, Kyoto University.\\
Kitashirakawa Oiwakecho Sakyo-ku, Kyoto 606-8502, Japan}
\date{\today}
\begin{abstract}
We obtained steady solutions to the two-dimensional Boussinesq
approximation equations without mean temperature gradient. This system 
is referred to as free convection in this paper.
Under an external flow described by the stream function 
$\mPsi = - Ayf(x)$, two types of steady solutions are found depending on 
the boundary conditions. 
One is kept steady by the balance between the strain of 
$\mPsi$ and the diffusion. The solution is similar to the Burgers vortex layer solution.
The other is done by the balance between vorticity induced by the buoyancy
and vorticity flux caused by the external flow.
Detailed argument on these two balances is presented for $f(x) = x$. Then
two examples other than $f(x) = x$ are shown to have either of the two balancing
mechanism. We discuss the relation between these solutions and
long-lived fine scale coherent structures observed  in direct numerical
simulations of two-dimensional free convection turbulence. 
\end{abstract}
\pacs{47.27.-i, 47.27.Te, 47.32.Cc}

\maketitle

\section{Introduction}
\label{s:intro}
Rayleigh--B\'{e}nard convection is very rich in nature. It has been a
source of inspiration in studies of universality in the route to
turbulence \cite{cvitanovic89}. At the same time, there have been
extensive researches both experimental and theoretical on turbulent
states of convection \cite{siggia94}.

Recently, convection turbulence has attracted
attention as a good example of active scalar systems. One interesting
observation is scaling behavior of structure functions of temperature, active scalar.
In this context, a two-dimensional Boussinesq convection model has been studied
by direct numerical simulations (DNS) in a doubly periodic square domain.
The results indicate that the scaling exponents of the temperature
structure functions saturate as the order goes to infinity \cite{cmv01, bhs01}.
It has been pointed out that coherent structures like plumes or sharp
interfaces (shocks) between hot and cold regions can play an important
role in this behavior.
Moreover it is shown phenomenologically \cite{l91} and numerically \cite{ts94} that the
temperature variance (entropy \cite{l91}) cascades from larger scales to smaller
scales in the turbulent two-dimensional Boussinesq system (referred to as
 two-dimensional free convection, 2DFC) when the
turbulent state is maintained by a large scale temperature forcing.
However, the relation between the entropy cascade and the dynamics of
coherent structures, if any at all, is still far from understood.

In the three-dimensional Navier-Stokes (3DNS) turbulence, vortical fine
coherent structures, so-called worms, are observed. These worms are well known
to be  approximated  locally by the Burgers vortex tube. In this sense, 
2DFC turbulence is quite similar to 3DNS turbulence, although the former
is more feasible than the latter. 
We expect that the understanding of the former may help us understand the
latter and also elucidate the universal characteristics 
of turbulence. 

The motivation of this paper is to explain some properties of the long-lived
coherent structures observed in DNS in terms of  a class of steady
solutions of 2DFC.
For this we search for steady solutions to the  incompressible
two-dimensional Boussinesq approximation equations without mean
temperature gradient (2DFC):
\begin{eqnarray}
\partial_t T
 + \left(\vec{u} \cdot \nabla \right) T
 &=& \kappa \triangle T,
\label{e:temp0} \\
\partial_t \vec{u}
 + \left(\vec{u} \cdot \nabla \right) \vec{u}
 &=& - {\nabla p \over \rho_0} + \alpha g T \hat{\vec{e}} + \nu \triangle \vec{u},
\label{e:velocity0} \\
\nabla \cdot \vec{u} &=& 0.
\label{e:incomp0} 
\end{eqnarray}
Here $\kappa$, $\rho_0$, $\alpha$, $g$, and $\nu$ are
the molecular diffusivity, the mean density of the fluid (we take $\rho_0$ to
be unity for simplicity), the thermal expansion coefficient, the
gravitational acceleration and the kinematic viscosity, respectively.
The vector $\hat{\vec{e}}$ is the unit vector in the direction opposite to
the gravity. 
In DNS of free convection turbulence, to keep the system statistically 
stationary,  a large-scale forcing term is added to Eq.(\ref{e:temp0}),
but the details are not mentioned here.
 
In the turbulent temperature field,
long-lived sharp interfaces between hot and cold regions, i.e. 
shock fronts are often formed and regarded as  coherent structures.
Thus coherent structures  are highly elongated; 
The typical length  is the order of the integral scale
and the typical width the order of ten times of Kolmogorov
dissipation length scale. 
We expect that steady solutions to
Eqs. \eqref{e:temp0}-\eqref{e:incomp0} have common properties with the
coherent structures formed spontaneously in DNS. 

To deal with temperature shocks more directly, we use the following 
vector quantity \cite{ky84, c94}:
\begin{equation}
 \vec{\chi} = \left( \partial_y T, -\partial_x T \right).
\label{e:defchi}  
\end{equation}
We call $\vec{\chi}$ T-vorticity. T-vorticity obeys the following evolution 
equation similar to the three-dimensional vorticity equation:
\begin{equation}
 \partial_t \vec{\chi}
  + \left(\vec{u} \cdot \nabla \right)\vec{\chi}
  = \left(\vec{\chi} \cdot \nabla \right) \vec{u} + \kappa \triangle \vec{\chi}.
\label{e:chi0}  
\end{equation}
Thus T-vorticity is expected to play an essential role in turbulence 
like vorticity in 3DNS.
In contrast with T-vorticity, the governing equation for
 the vorticity $\omega(x, y, t)$ in 2DFC can be written
\begin{equation}
  \partial_t \omega
  + \left(\vec{u} \cdot \nabla \right)\omega
  =  \alpha g \{\nabla \times (T \hat{\vec{e}})\}_z + \nu \triangle \omega.
\label{e:omega0}  
\end{equation}
Here $\{\cdot\}_z$ denotes taking the $z$-component.
Note that because of the buoyancy term, i.e. the first term in the 
r.h.s. of Eq.\eqref{e:omega0}, vorticity is not a conservative quantity even for
$\nu = 0$. 
For simplicity, we assume  $\nu = \kappa$ , i.e. Prandtl number 
$Pr = \nu / \kappa = 1$ here. 

In the following sections, we look for steady solutions 
to Eqs. \eqref{e:chi0} and
\eqref{e:omega0},  where the flow is decomposed 
into the external part given by a stream function $\mPsi = - Ayf(x)$
 and a response to it. 
In section II, we deal with the case $f(x)=x$, i.e. a stagnation 
flow. 
In Section III,  extended  stagnation flows, $f(x)\ne x$, are considered,
where we show two examples of $f(x)$.  The realizability of the 
extended external flow is discussed.
Concluding remarks are made in \secref{s:cr}.

\section{Burgers vortex layer type solutions}
\label{s:burgers}
\begin{figure}
\iffigs
\includegraphics[scale = 1.8]{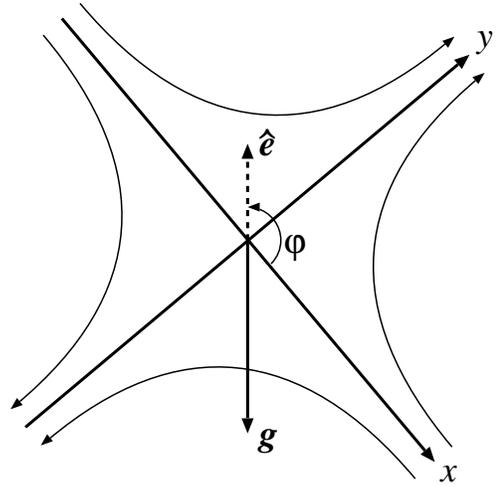} 
\else\drawing 65 10 {coordinate setting}
\fi
\caption{\label{f:coord}The coordinate system fixed on the external
 stagnation flow. The direction of the gravity vector $\vec{g}$ is
 expressed via the angle $\varphi$.}
\end{figure}

We assume that the system \eqref{e:chi0} and \eqref{e:omega0}
is exposed to a stagnation flow
\begin{equation}
 \mPsi(x, y) = - A x y, 
\label{e:burg-ext}  
\end{equation}
where $A$ is a positive constant so that
 we take the directions of contraction and expansion
as $x$-axis and $y$-axis, respectively. The angle between $x$-axis and 
$\hat{\vec{e}}$ is also a  parameter and denoted by $\varphi$ as shown in
\figref{f:coord}. Then the unit vector $\hat{\vec{e}}$ in Eq. \eqref{e:temp0}
is given in this coordinate as $\hat{\vec{e}} = (\cos \varphi, \sin\varphi)$.
We use this coordinate throughout this paper.
We further assume the following forms of the temperature and velocity
fields:
\begin{eqnarray}
 T(x, y, t) &=& \theta(x),
\label{e:burg-tempdep} \\
 \vec{u}(x, y, t) &=& (\partial_y \mPsi(x, y),-\partial_x \mPsi(x, y))
 + (0, v(x)) \nonumber \\
                  &=& (-Ax, Ay + v(x)).
\label{e:burg-veldep}  
\end{eqnarray}
In terms of $\theta$ and $v$, T-vorticity and vorticity are expressed as
\begin{eqnarray}
 \vec{\chi}(x, y, t) &=& (0, -\partial_x \theta(x)) \equiv (0, \chiy(x)),
\label{e:burg-chidep} \\
 \omega(x, y, t) &=& \omega(x) = \frac{dv(x)}{dx}.
\label{e:burg-omegadep}  
\end{eqnarray}
Consequently, Eqs. \eqref{e:chi0} and
\eqref{e:omega0} are reduced to the following ordinary differential equations:
\begin{eqnarray}
&& {d \over dx} \left( {\kappa \over A} {d\chiy \over dx} + x\chiy \right) = 0,
\label{e:steady_eq_chi} \\
&& {\nu \over A} {d^2 \omega \over dx^2} + x{d\omega \over dx} = \chi_y {\alpha g \sin\varphi \over A}. 
\label{e:steady_eq_omega}  
\end{eqnarray}
After integrating once,  Eq.\eqref{e:steady_eq_chi} can be written
\begin{equation}
 {\kappa \over A} {d\chiy \over dx} + x \chiy =C.
\label{e:burg_2cases} 
\end{equation}
Here $C$ is an integration constant.

Let us consider the
general boundary conditions at the origin:
\begin{eqnarray}
 \chiy(x = 0)   &=& \chi_0, \label{e:bc_chi0}\\
 \chiy'(x = 0)  &=& \chi_0', \label{e:bc_dchi0}\\
 \omega(x = 0)  &=& \omega_0, \label{e:bc_omega0}\\
 \omega'(x = 0) &=& \omega_0' \label{e:bc_domega0}, 
\end{eqnarray}
where $\chiy'(x)$ and $\omega'(x)$ denote the derivatives of T-vorticity and
vorticity respectively; $\chi_0, \chi_0', \omega_0$ and $\omega'_0$ are constants.
These conditions yield the relation $C = \kappa \chi_0' / A$.
With $\kappa = \nu$ the solutions to Eqs.\eqref{e:steady_eq_chi} and
\eqref{e:steady_eq_omega} are obtained
\begin{eqnarray}
 \chiy(x) &=& \chi_0 ~ e^{-{A \over 2\kappa} x^2}
          + \chi_0'~ e^{-{A \over 2\kappa} x^2}
	    \int_0^x  e^{{A \over 2\kappa}\xi^2} d\xi, \label{e:gen_sol_chi}
 \label{e:sol_chi_gen}	    \\
 \omega(x)&=& \omega_0 + \omega_0' \int_0^{x} e^{-{A \over \kappa}\xi^2} d\xi
           + \chi_0 {\alpha g \sin\varphi \over A} ( 1 - e^{-{A \over 2\kappa}x^2}) \nonumber \\
	  && + \chi_0'{\alpha g \sin\varphi \over A}
	        \int_0^{x} d\zeta e^{-{A \over 2\kappa}\zeta^2}
		 \int_0^{\zeta} d\mu
		 \int_0^{\mu} d\xi~ e^{{A \over 2\kappa} \xi^2}. \nonumber \\
 \label{e:sol_omega_gen}
\end{eqnarray}
Here only the last two terms of Eq.\eqref{e:sol_omega_gen}
are the buoyancy generated vorticity. The first two are homogeneous
solutions, i.e., the solutions to the vorticity equation
\eqref{e:steady_eq_omega} setting the r.h.s. (the buoyancy force) zero.
When $\chi_0' = 0$, the solution \eqref{e:gen_sol_chi} which is the same
as the Burgers vortex layer, takes finite value of temperature
at $x = \pm \infty$. We call this solution finite temperature (FT) type solution.
In contrast, when $\chi_0' \ne 0$, the temperature corresponding to the
solution \eqref{e:gen_sol_chi} diverges as $x \to \pm \infty$.
We then call this solution infinite temperature (IT) type solution. We treat
the two cases separately in the following  subsections.

\subsection{Finite temperature type solutions}
\begin{figure}
\iffigs
\includegraphics[width=8.5cm]{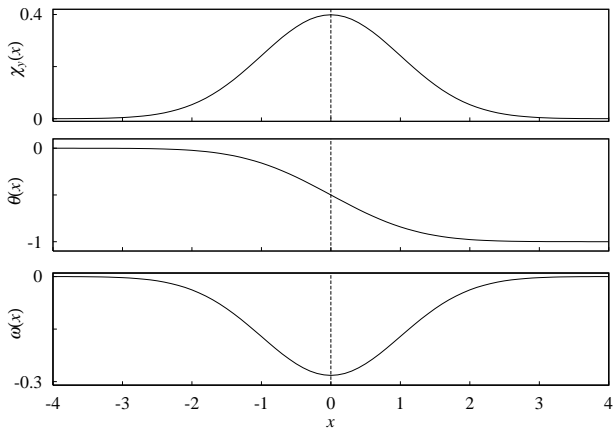} 
\else\drawing 65 10 {mono burg}
\fi
\caption{\label{f:monoburg}Finite temperature (FT) solutions:
the T-vorticity Eq.\eqref{e:burg-sol-chi},  the corresponding temperature
 $\theta(x)$ and the vorticity Eq. \eqref{e:burg-sol-omega}.
Here we set the parameters as
 $\Theta_0 = 1, A = 1, \kappa = 1, \alpha g = 1$ and $\varphi = \pi / 4$.}
\end{figure}

Let us consider the following boundary conditions:
\begin{eqnarray}
 \chiy(x) &\to& 0  \quad (x \to \pm \infty),
\label{e:burg-bcchi}  \\
 \omega(x)  &\to& 0  \quad (x \to \pm \infty).
\label{e:burg-bcomega}  
\end{eqnarray}
These conditions \eqref{e:burg-bcchi} and \eqref{e:burg-bcomega} are
equivalent to $\chi_0' = 0$ and $\omega_0' = 0$.
The general solutions \eqref{e:sol_chi_gen} and \eqref{e:sol_omega_gen}
are then reduced to Burgers vortex layer type solution:
\begin{eqnarray}
\chiy(x) &=& 
  \mTheta_0 \sqrt{A \over 2\pi \kappa}
  \exp \left( - {A \over 2 \kappa} x^2 \right),
\label{e:burg-sol-chi} \\
 \omega(x) &=& -{\alpha g \sin\varphi \over A}
            \mTheta_0 \sqrt{A \over 2 \pi \kappa}
	    \exp\left( - {A \over 2\kappa} x^2 \right) \nonumber \\
           &=& -{\alpha g \sin\varphi \over A} \chiy(x),
\label{e:burg-sol-omega}	    
\end{eqnarray}
where $\mTheta_0 = \int_{-\infty}^{\infty} \chiy(\xi, t=0) ~d\xi =
\theta(-\infty, 0) - \theta(\infty, 0)$.
In the Burgers vortex layer, this temperature difference
corresponds to the circulation of the vortex layer.
In \figref{f:monoburg} we plot the solutions; T-vorticity \eqref{e:burg-sol-chi},
and the corresponding temperature $\theta(x)$ and vorticity \eqref{e:burg-sol-omega}.
As mentioned before, the temperature takes finite value at $x = \pm \infty$.
Since the temperature $\theta(x)$ is a monotonic function,
$\chi_y(x)$ contains a single bump. We call this solution Burgers
T-vortex layer solution.

Now we make several remarks on the FT solutions \eqref{e:burg-sol-chi} and
\eqref{e:burg-sol-omega}.
First, 
we note that $\omega(x)$ is proportional to $\chiy(x)$ and
$\sin\varphi$. 
That is, $\chiy(x)$ is independent of the direction of the gravity, but 
$\omega(x)$ is not. This is because the former is maintained by the 
balance between the 
imposed strain  and the diffusion of  T-vorticity like the Burgers 
vortex layer solution  to the 3DNS equations and the latter does 
by the balance between the diffusion of vorticity and the buoyancy
induced by the temperature difference across 
the Burgers T-vortex layer.  Therefore vorticity is maximized when the
direction of the T-vortex layer coincides  with that of  the gravity.

Secondly,
it is possible to estimate a characteristic time of
the relaxation to the steady state \eqref{e:burg-sol-chi}
after the temperature difference $\mTheta_0$ is fixed.
Under the stagnation flow \eqref{e:burg-veldep},  the time-dependent 
solution to Eq. \eqref{e:chi0} is obtained \cite{l82} as
\begin{eqnarray}
 \chiy(x, t) &=&
  \left( 4 \pi \kappa { 1 - e^{-2At} \over 2A}\right)^{-1/2} \nonumber \\
  &\times& \int_{-\infty}^{\infty} \chiy(\xi, t = 0) 
  \exp\left[ - { ( x - e^{-At} \xi)^2 \over 4 \kappa {1 - e^{-2At} \over 2A}}\right] d\xi \nonumber \\
 &\to&
  \mTheta_0 \sqrt{A \over 2\pi \kappa}
  \exp \left( - {A \over 2 \kappa} x^2 \right)
 \quad \text{as}~ t \to \infty.
 \label{e:burg_time-dep-chi}
\end{eqnarray}
The characteristic time for the relaxation is thus estimated as $1/A$.

Thirdly,
 the characteristic width of the bump of
T-vorticity and vorticity  is $\sqrt{2\kappa / A}$. 
This agrees with an observation about structures of a passive
scalar advected by a two-dimensional synthetic random velocity 
field with a finite
correlation time in Ref.~\cite{hs94}. (Of course, vorticity is not
considered in that case.)
Although  a passive scalar with non-zero mean gradient was considered,
the solution \eqref{e:burg-sol-chi} may describe locally 
coherent structures of the passive scalar because of the following features:
The width of those was found to scale well with $\sqrt{\kappa / s}$, 
where $s$ was a root mean square of the
rate of strain calculated from the low-path filtered velocity.
They also showed a stagnation flow accompanied by a coherent
structure in Fig. 7 of Ref.~\cite{hs94}.

Finally,
concerning the relevance of the FT solutions
\eqref{e:burg-sol-chi} and \eqref{e:burg-sol-omega} to
the coherent structures observed in DNS of free
convection turbulence, we refer the reader to Ref.~\cite{tm01}. 
Here suffice it to say that the solutions \eqref{e:burg-sol-chi} 
and \eqref{e:burg-sol-omega} can give a good local approximation of 
the coherent structures in DNS, 
where  we compared  $A$ with the (positive) eigen-value of the 
velocity-gradient tensor although the latter is defined only locally.

However, it is well-known that, in the limit of $\kappa \to 0$, 
the temperature dissipation rate per unit length along $y$-axis,
$\epsilon_\theta$, of the Burgers (FT type) solution \eqref{e:burg-sol-chi}
vanishes. That is, $\epsilon_\theta = \int_{-\infty}^{\infty} \kappa
\chiy(x)^2~dx \propto \sqrt{\kappa A} \to 0$ as $\kappa \to 0$.
It suggests that this Burgers layer type solution is irrelevant in
this limit.
In DNS, however, the situation can be different because we have to
consider feedback to $A$ (the eigen-value of
the velocity-gradient tensor) from other ingredients of the flow.
Indeed, the probability distribution functions of $A(x, y)$ obtained in
DNS of various $\kappa$'s are observed to fall into a unique curve under
a suitable normalization. In other words, $A(x, y)$ field
may have some nontrivial dependency on $\kappa$.
Thus it is conjectured that the dissipation rate of the FT type
solution \eqref{e:burg-sol-chi}, $\sqrt{\kappa A}$, would take
non-zero value in the limit of vanishing diffusivity in real flows.
Therefore the relevance of the Burgers vortex layer type solutions 
might not be ruled out. The detail of the observation on $A(x, y)$ field 
in DNS will be reported elsewhere.

\subsection{Infinite temperature type solutions}
\begin{figure}
\iffigs
\includegraphics[width = 8.5cm]{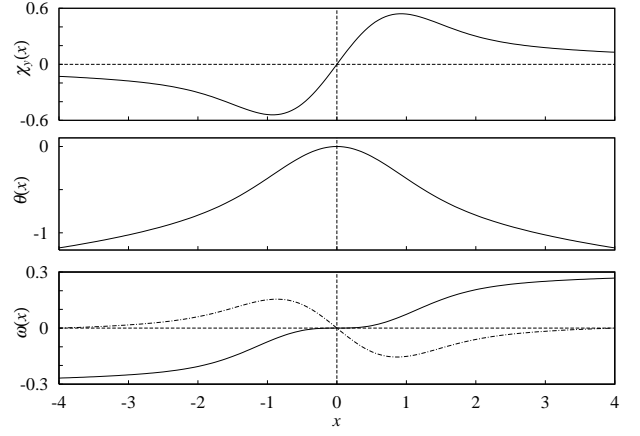} 
\else\drawing 65 10 {nonmono-burg}
\fi
 \caption{\label{f:non-monoburg}Infinite temperature (IT) solutions:
 the T-vorticity Eq.\eqref{e:sol-dbs-chi},the
 corresponding temperature $\theta(x)$ and the vorticity
 Eq.\eqref{e:sol-dbs-omega}.
 Here we set the parameters as  $A = 1, \kappa = 1, \alpha g = 1, \varphi = \pi / 4, \chi_0' = 1,
  \theta(x = 0) = 0, \omega'_0 = 0$ and $\kappa = 1$. The
 dash-dotted line in the vorticity figure is a localized vorticity
 solution Eq.\eqref{e:canceled_vorticity} obtained by setting $\omega_0' = -0.23$.}
\end{figure}

In this subsection, we consider the case where the temperature slowly
diverges as $x \to \pm\infty$, which we call infinite temperature
type solutions. In terms of the boundary conditions at
the origin, it corresponds to
\begin{eqnarray}
 \chiy(x = 0)   &=& 0,       \label{e:itd_bc_chi}\\
 \chiy'(x = 0)  &=& \chi_0', \label{e:itd_bc_dchi}\\
 \omega(x = 0)  &=& 0,       \label{e:itd_bc_omega}\\
 \omega'(x = 0) &=& 0,       \label{e:itd_bc_domega}
\end{eqnarray}
where $\chi_0' \ne 0$. We set $\chi_0=0$ to examine an IT type solution
solely.
The general solutions \eqref{e:sol_chi_gen} and \eqref{e:sol_omega_gen}
then take the form:
\begin{eqnarray}
 \chi_y(x) &=&   \chi_0' e^{-{A\over 2\kappa}x^2} \int_0^{x} e^{{A\over2\kappa}\xi^2} d\xi,
 \label{e:sol-dbs-chi} \\
 \omega(x) &=& 
               \chi_0'{{\alpha g \sin\varphi} \over A}
	       \int_{0}^{x} \! d\zeta ~e^{-{A \over 2 \kappa} \zeta^2}
	       \int_{0}^{\zeta} \! d\mu
	       \int_{0}^{\mu} \! d\xi  ~e^{{A \over 2 \kappa}\xi^2}.
\label{e:sol-dbs-omega}	       
\end{eqnarray}
 The boundary conditions for the vorticity \eqref{e:itd_bc_omega} and
 \eqref{e:itd_bc_domega} are chosen to involve only the buoyancy effect
 on vorticity since we focus on the vortical structure induced
 by the buoyancy. It should be noted that the homogeneous solutions of Eq.\eqref{e:steady_eq_omega}
 are neglected.
We plot the IT solutions:T-vorticity Eq.\eqref{e:sol-dbs-chi}
,the corresponding temperature $\theta(x)$ and vorticity
 Eq.\eqref{e:sol-dbs-omega} in \figref{f:non-monoburg}.

It should be noted that unlike the Burgers type solutions
T-vorticity and vorticity of IT type solutions 
take the same sign around $\chi$-bumps.
This behavior seems to contradict physical intuition that hot fluid rises
and then vorticity with the sign opposite to that of T-vorticity
is induced. 
\begin{figure}
\iffigs
\includegraphics[width = 8.5cm]{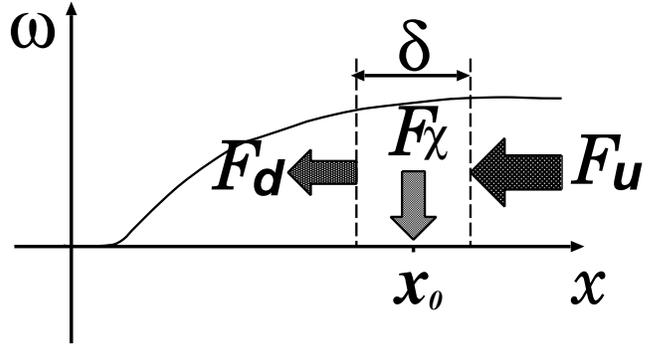} 
\else\drawing 65 10 {balance}
\fi
 \caption{\label{f:balance}Schematic view of vorticity balance in a
 region with the width $\delta$ centered at $x_0$.
 $F_u=A(x_0+\delta/2)\omega(x_0+\delta/2)$ and
 $F_d=-A(x_0-\delta/2)\omega(x_0-\delta/2)$ are vorticity fluxes in $x$ by a
 stagnation flow.  $F_{\chi}=\delta\chiy(x_0){\alpha g \sin \varphi \over A}$ is
 vorticity induced by the buoyancy.  
}
\end{figure}
However, vorticity is maintained  by  the balance between 
the buoyancy and not only the diffusion of vorticity  but  the vorticity
 transfer by a stagnation flow. In fact, if vorticity takes non-zero 
value with the same sign of T-vorticity, say positive, at infinity,
vorticity flux $Ax\omega$ through a region with the width $\delta$ 
centered  at some point ,e.g. $x_0$,  always increases  vorticity in the 
region. The substantial increase in vorticity $Q_{\omega}$ is given by
\begin{eqnarray}
Q_{\omega} &=& F_u -F_d \nonumber \\
 &=&  A \left(x_0+{\delta \over 2}\right) \omega\left(x_0+{\delta \over 2}\right)
    - A \left(x_0-{\delta \over 2}\right) \omega\left(x_0-{\delta \over 2}\right) \nonumber \\
 &=&  A\delta\omega(x_0) + A \delta  x_0 \omega'(x_0),
\label{e:balance}
\end{eqnarray}
where the first term in the r.h.s balances with the vorticity flux in $y$ and 
the second term does with $F_{\chi}= \delta \chiy(x_0) \alpha g \sin\varphi/A$ if $\omega''(x_0)$ is 
negligible in Eq.\eqref{e:steady_eq_omega} as shown in \figref{f:balance}. That is, vorticity induced by 
the buoyancy cancels out the excess of the vorticity flux. On the other hand,
if the sign of vorticity around infinity  is oppose to that of T-vorticity,
the excess of the vorticity flux cannot be canceled out by the vorticity 
induced by the buoyancy so that the vorticity is not kept steady.

When vorticity is localized around the origin, i.e, vanishes rapidly
as $x$ goes to infinity, the vorticity transfer is turned off and 
the vorticity induced by the buoyancy works as 
the vorticity input which balances with the diffusion of vorticity. 
This means that to keep  the vorticity steady,  $\omega(\infty)$ should 
be null or take the same sign as that of the T-vorticity. Even in the IT
case, using the homogeneous solution to Eq.\eqref{e:steady_eq_omega}, vorticity can be localized around the origin.
In \figref{f:non-monoburg} with dash-dotted line,
we show this localized vorticity, which consists of the buoyancy-induced
vorticity \eqref{e:sol-dbs-omega} and the homogeneous solution,
\begin{eqnarray}
 \tomega(x)&=&
  \omega_0' \int_0^{x} e^{-{A \over \kappa}\xi^2} d\xi \nonumber \\
 &+&
  \chi_0'{{\alpha g \sin\varphi} \over A}
  \int_{0}^{x} \! d\zeta ~e^{-{A \over 2 \kappa} \zeta^2}
  \int_{0}^{\zeta} \! d\mu
  \int_{0}^{\mu} \! d\xi  ~e^{{A \over 2 \kappa}\xi^2}.
\label{e:canceled_vorticity}  
\end{eqnarray}
Here the value for the parameter $\omega_0'$ is chosen to cancel out the
uniform vorticity of the solution \eqref{e:sol-dbs-omega}. The form of this localized response
vorticity matches the intuition that hot fluid goes up and cold fluid down.
Such a localized vorticity structure associated with
T-vorticity structure is one of the striking features of active
scalar turbulence.

For any non-vanishing value of $\chi_0'$,
$\chi_y(x)$ is an odd function having a double bump and
consequently the temperature $\theta(x) =
\theta(x=0) - \int_0^x \chiy(\xi) d\xi$ forms a ridge.
Such temperature ridges are often observed in DNS.
However, far away from the origin, temperature diverges logarithmically.
\cite{bo78}. 
We think that the divergence at infinite distances is not necessarily 
pathological, since we are interested only in the local behavior of the
structures (around the origin). Furthermore, this divergence prevents the 
strain of the external flow from squashing the temperature ridge.
Like the FT type, the typical width of the vorticity structure is estimated as
 $\sqrt{\kappa / A}$.
It is noted that this IT solution \eqref{e:sol-dbs-chi} 
is also the Burgers vortex layer solution to the three-dimensional 
barotoropic vorticity equation.
 (Recall that the governing equation \eqref{e:chi0} for $\vec{\chi}$ 
is the same as that of three-dimensional barotropic vorticity.)

\section{extension}
\label{s:ayf}
In this section, we consider the following external flow:
\begin{eqnarray}
 \mPsi(x, y) = - Ayf(x),
\label{e:ext1-Psi}  
\end{eqnarray}
where $f(x)$ is a function satisfying a condition mentioned later.
When $f(x)=x$, this flow becomes a stagnation flow.
A major difference from the $f(x) = x$ cases is that the external flow \eqref{e:ext1-Psi} has 
in general non-zero vorticity, 
\begin{equation}
\mOmega(x, y) \equiv - \triangle \mPsi = Ayf''(x).
\label{e:ext1-Omega}
\end{equation}
Thus $\mOmega$ should satisfy the vorticity equation.
We will discuss the realizability of \eqref{e:ext1-Psi} in detail later
in this section. 

Under the extended stagnation flow \eqref{e:ext1-Psi},
 we further assume functional forms of temperature and velocity as follows:
\begin{eqnarray}
 T(x, y, t) &=& \theta(x),
 \label{e:ext1-t-dep} \\
 \vec{u}(x, y, t) &=& (\partial_y \mPsi, -\partial_x \mPsi) + (0, v(x)) \nonumber \\
                  &=& ( -Af(x), Ayf'(x) + v(x)).
 \label{e:ext1-vel-dep}
\end{eqnarray}
Then $\vec{\chi}$ and the total vorticity $\omega$ are given by 
\begin{eqnarray}
 \vec{\chi}(x) &=& (0, -\theta'(x)) \equiv (0, \chiy(x)),
\label{e:ext1-chi-dep} \\
 \omega(x, y) &=& - \triangle \mPsi + v'(x) \equiv \mOmega(x, y) + \tomega(x).
\label{e:ext1-omega-dep} 
\end{eqnarray}
We call $\tomega$ response vorticity.
The velocity field Eq. \eqref{e:ext1-vel-dep} is uniquely decomposed into the
$y$-dependent and the $y$-independent parts.

Imposing the general boundary conditions
\begin{eqnarray}
 \chiy(x = 0)   &=& \chi_0,       \label{e:ext_bc_chi}\\
 \chiy'(x = 0)  &=& \chi_0', \label{e:ext_bc_dchi}\\
 \tomega(x = 0)  &=& \tomega_0,       \label{e:ext_bc_omega}\\
 \tomega'(x = 0) &=& \tomega_0',       \label{e:ext_bc_domega}
\end{eqnarray}
steady solutions to Eqs.\eqref{e:chi0} and \eqref{e:omega0} are easily
obtained for any $f(x)$ as follows:
\begin{eqnarray}
 \chiy(x) &=&
  \chi_0 e^{-{A \over \kappa} \int_0^{x} f(\lambda) ~d\lambda} \nonumber \\
 && + \chi_0'  e^{-{A \over \kappa} \int_0^{x} f(\lambda) ~d\lambda}
       \int_0^{x} e^{{A \over \kappa} \int_0^{\xi} f(\lambda) ~d\lambda}
	  ~d\xi,  
\label{e:ext1-sol-chi} \\
 \tomega(x) &=& \tomega_0 + \tomega_0' \int^x_0
  e^{ - {A \over \nu} \int_0^{\zeta} f(\lambda) d\lambda} ~d\zeta \nonumber \\
+&&\hspace{-18pt}{\alpha g \sin \varphi \over A}
\int^x_0 d\zeta e^{ - {A \over \nu} \int^{\zeta}_0 f(\lambda) d\lambda}
\int^{\zeta}_0 \chiy(\mu) e^{{A \over \nu}\int^{\mu}_0 f(\lambda) d\lambda} d\mu \nonumber \\
 &=& \tomega_0 + \tomega_0' \int^x_0 e^{ - {A \over \nu} \int^{\zeta} f(\lambda) d\lambda} ~d\zeta \nonumber \\
 && + \chi_0 {\alpha g \sin \varphi \over A}
      \int_0^x \zeta e^{ - {A \over \nu} \int^{\zeta} f(\lambda) d\lambda} d\zeta \nonumber \\
 && + \chi_0' {\alpha g \sin \varphi \over A} 
      \int^{x}_0 d\zeta e^{ - {A \over \nu} \int^{\zeta} f(\lambda) d\lambda} \nonumber \\
 && \hspace{12pt}\times      
      \int_0^{\zeta} d\mu \int_0^{\mu} d\xi  e^{{A \over \nu}\int^{\mu}_0 f(\lambda) d\lambda} d\mu
\label{e:ext1-sol-tomega}     
\end{eqnarray}
The last two terms of Eq.\eqref{e:ext1-sol-tomega} represent the vorticity
induced by the buoyancy.

Now we discuss the realizability of the external flow \eqref{e:ext1-Psi}.
If we claim that Eq.~\eqref{e:ext1-Omega} is a steady solution of the
barotropic two-dimensional vorticity equation
\begin{equation}
(\vec{U}\cdot\nabla) \mOmega = \nu \triangle \mOmega,
\label{e:eq-Omega} 
\end{equation}
where $\vec{U} = (\partial_y \mPsi, -\partial_x \mPsi)$.
Then the equation for $f(x)$ is
\begin{eqnarray}
 {\nu \over A}f^{(4)}(x) + f(x) f^{(3)}(x) - f'(x)f''(x) = 0,
\label{e:ext1-eqf}  
\end{eqnarray}
where $f^{(j)}(x)$ denotes the $j$-th order derivative.
Equation \eqref{e:ext1-eqf} has been
studied by many authors as a model of a flow near a rigid
wall \cite{b67}.
In Ref.~\cite{co00} the authors, who were interested in the finite
time blow-up of the solution, dealt with the time-dependent stream
function $\mPsi = - Ayf(x, t)$ in a bounded domain and discussed
asymptotic behavior ($t \to \infty$) of the solution to the equation for $f(x, t)$,
the Proudman-Johnson equation \cite{pj61}.
They showed that every solution decays to zero as $t \to \infty$ with
homogeneous boundary conditions.

Here instead of  solving Eq. \eqref{e:ext1-eqf}, 
we examine the realizability of the external field \eqref{e:ext1-Psi}.
 We do not think that the flow \eqref{e:ext1-Psi} should  be
an exact solution of the vorticity equation \eqref{e:eq-Omega}.
Rather, we assume that the flow \eqref{e:ext1-Psi} is formed by a
larger scale flow  in  the turbulence field.
That is, from a practical point of view, 
a steady solution subjected to an external flow 
is no more than  a local model of coherent structures in 
the turbulent field, so that the external flow need not satisfy
 Eq.\eqref{e:eq-Omega}.
Thus it may be possible to add an forcing term describing 
an effect of the larger scale motions to  Eq.\eqref{e:eq-Omega}.
Denoting the forcing term by $F(x,y)$, 
the equation for $\mOmega$ (or $f(x)$) can be rewritten 
\begin{eqnarray}
 (\vec{U} \cdot \nabla) \mOmega = \nu \triangle \mOmega + F.
\label{e:ext1-eq-Omega}  
\end{eqnarray}
Then the steady-state equation for the total vorticity
 \eqref{e:ext1-omega-dep} can be written 
\begin{equation}
 (\vec{u} \cdot \nabla)(\mOmega + \tomega) =
  \chiy(x) \alpha g \sin \varphi 
  + \nu \triangle (\mOmega + \tomega) + F,
\label{e:ext1-eq-tomega}  
\end{equation}
which yields the equation for the response vorticity $\tomega$
\begin{equation}
 \frac{\nu}{A} \frac{d^2 \tomega}{dx^2} + f(x) \frac{d\tomega}{dx} = \chi_y(x) \frac{\alpha g \sin \varphi}{A},
\label{e:ext1-eq-res-omega}  
\end{equation}
from which the solution \eqref{e:ext1-sol-tomega} is obtained.

By substituting $f(x) = x$, it is easily checked that the 
solutions \eqref{e:ext1-sol-chi} and \eqref{e:ext1-sol-tomega} contain
Burgers vortex layer type solutions. 

\begin{figure}[htbp]
\iffigs
\includegraphics[scale=0.85]{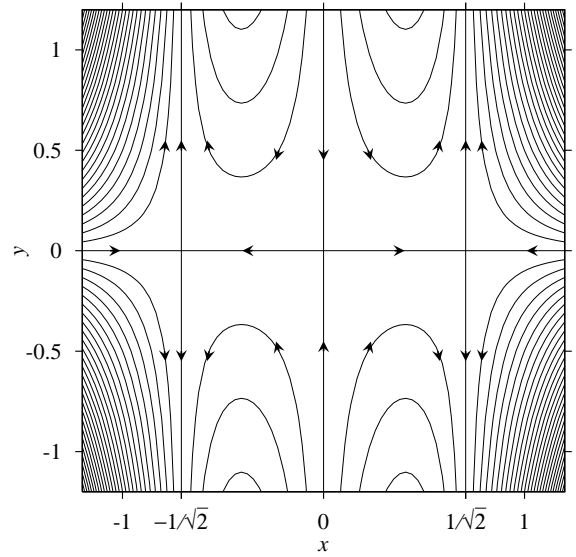} 
\else\drawing 65 10 {ext1_field}
\fi
\caption{\label{f:ext1_stream} Stream lines of an external flow $\mPsi(x) = -y(4 x^3 - 2 x)$. The arrows denote the directions of the flow. Stagnation points are $(-1/\sqrt{2}, 0), (0, 0)$ and $(1/\sqrt{2}, 0)$.}
\end{figure}

In the following, we show two examples of $f(x)$. Each has
both FT and IT type solutions.
As the first example, we deal with an external flow
\begin{eqnarray}
 f(x) = 4 x^3 - 2 x, 
\label{e:ext1-specific-f}  
\end{eqnarray}
which itself has vorticity $\Omega(x, y) = 24Axy$.
The stream lines of the external flow is shown in
\figref{f:ext1_stream}.
Here we assume  an additional forcing $F(x, y)$ that keeps the 
external field \eqref{e:ext1-specific-f} steady.
( Equation \eqref{e:ext1-eq-Omega} prescribes the forcing as 
$F(x, y) = 192 A^2 x^3 y$)

\begin{figure}
\iffigs
\includegraphics[width = 8.5cm]{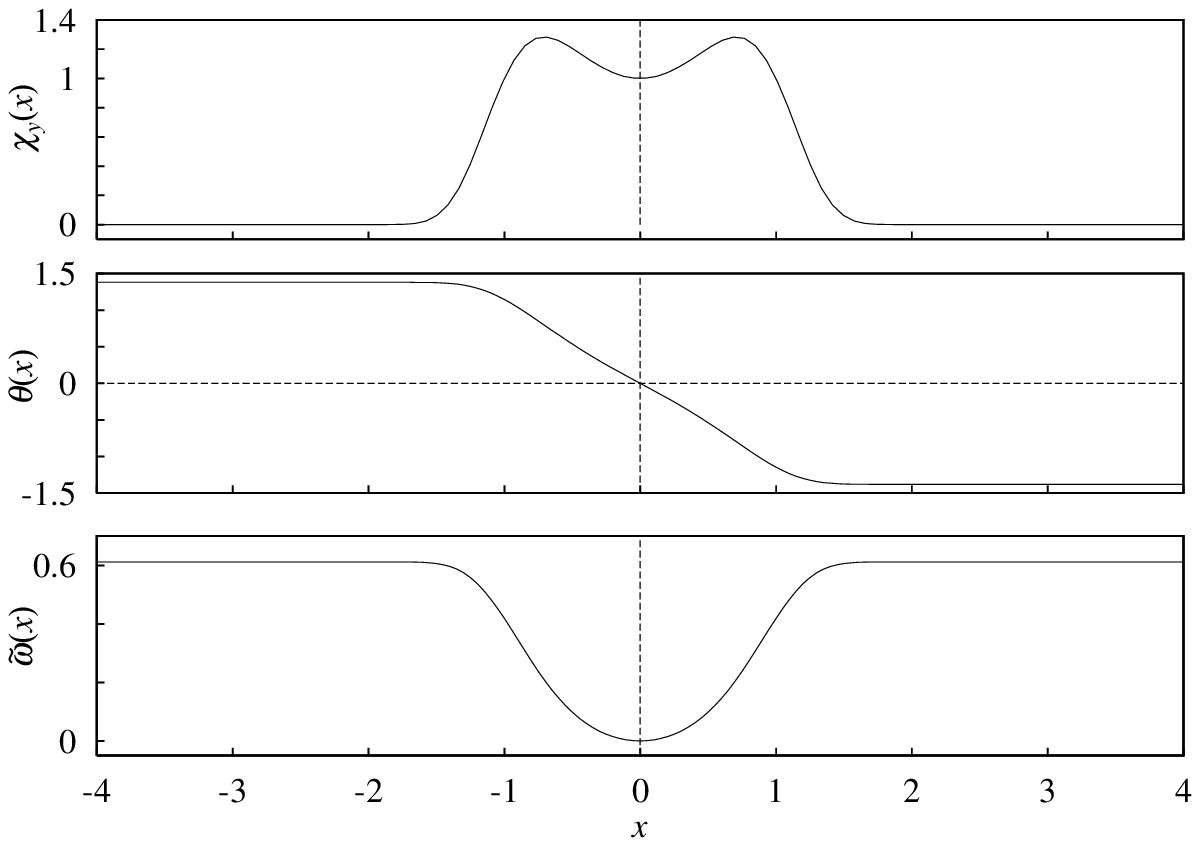} 
\else\drawing 65 10 {ext1, solutions, (a), (b), (c)}
\fi
\caption{\label{f:ftd_sol_1}FT type solutions for $\mPsi(x) = - Ay (4 x^3 - 2 x)$:
 the T-vorticity \eqref{e:ftd_sol_chi_1}, the corresponding temperature
 $\theta(x)$ and the response vorticity \eqref{e:ftd_sol_omega_1}. 
 Here we set the parameters as  $A = 1, \kappa = 1, \alpha g = 1, \varphi = \pi /4,
  \chi_0 = 1$ and $\theta(x = 0) = 0$.}
\end{figure}

For FT type solutions, we impose the boundary conditions
\begin{eqnarray}
\chiy(x = 0) = \chi_0 &\ne& 0,   \label{e:bc_ext_ftd_chi} \\
\chiy'(x = 0) = \chi_0' &=& 0,    \label{e:bc_ext_ftd_dchi} \\
\tomega(x = 0) = \tomega_0 &=& 0,   \label{e:bc_ext_ftd_tomega} \\
\tomega(x = 0)' = \tomega_0' &=& 0.  \label{e:bc_ext_ftd_dtomega}
\end{eqnarray}
The conditions for the vorticity Eqs.\eqref{e:bc_ext_ftd_tomega} and
\eqref{e:bc_ext_ftd_dtomega} are chosen again to reflect only the effect
of the buoyancy.
The solutions \eqref{e:ext1-sol-chi} and \eqref{e:ext1-sol-tomega}
are then reduced to 
\begin{eqnarray}
 \chiy(x) &=& \chi_0 e^{-{A \over \kappa} (x^4 - x^2)},
 \label{e:ftd_sol_chi_1}\\
 \tomega(x) &=& \chi_0 {\alpha g \sin\varphi \over A} \int_0^x \xi e^{-{A \over \kappa} (\xi^4 - \xi^2)} d\xi.
 \label{e:ftd_sol_omega_1}  
\end{eqnarray}
These solutions are plotted in \figref{f:ftd_sol_1}.

For IT type solutions, we use the boundary conditions
\begin{eqnarray}
\chiy(x = 0) = \chi_0     &=& 0,    \label{e:bc_ext_itd_chi} \\
\chiy'(x = 0) = \chi_0'    &\ne& 0,  \label{e:bc_ext_itd_dchi} \\
\tomega(x = 0) = \tomega_0  &=& 0,    \label{e:bc_ext_itd_tomega} \\
\tomega(x = 0) = \tomega_0' &=& 0.     \label{e:bc_ext_itd_dtomega}
\end{eqnarray}
Hence the solutions \eqref{e:ext1-sol-chi} and
\eqref{e:ext1-sol-tomega} is rewritten 
\begin{eqnarray}
 \chiy(x) &=& \chi_0' e^{-{A \over \kappa} (x^4 - x^2)}
              \int_{0}^x e^{{A \over \kappa} (\xi^4 - \xi^2)} ~d\xi,
\label{e:ext1-sp-sol-chi}  \\
\tomega(x) &=& \chi_0' {\alpha g \sin \varphi \over A}
 \int_{0}^{x} \! d\zeta ~e^{-{A \over \kappa} (\zeta^4 - \zeta^2)}
     \nonumber \\
 &&\times 
     \int_{0}^{\zeta} \! d\xi \int_{0}^{\xi} \! d\mu
     ~e^{{A \over \kappa}(\mu^4 - \mu^2)}
     . \nonumber \\
\label{e:ext1-sp-sol-omega}     
\end{eqnarray}
\begin{figure}
\iffigs
\includegraphics[width = 8.5cm]{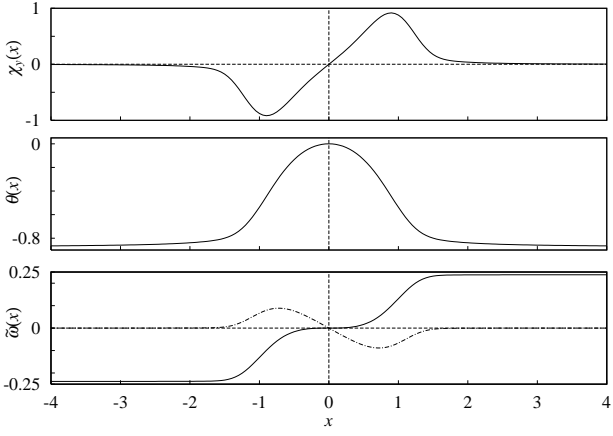} 
\else\drawing 65 10 {ext1, solutions, (a), (b), (c)}
\fi
\caption{\label{f:ext1-sp-sols}IT type solutions for $\mPsi(x) = - Ay (4 x^3 - 2 x)$:
 the T-vorticity Eq.\eqref{e:ext1-sp-sol-chi}, the corresponding
 temperature $\theta(x)$ and the response vorticity Eq.\eqref{e:ext1-sp-sol-omega}. 
 Here we set the parameters as  $A = 1, \kappa = 1, \alpha g = 1, \varphi = \pi /4,
  \chi'_0 = 1, \theta(x = 0) = 0,  
 $ and $\tomega'_0 = 0$. The temperature changes in a non-monotonic
 way. The dash-dotted line in the $\tomega$ figure is a localized response vorticity obtained by
 setting $\tomega'_0 = -0.172$.}
\end{figure}
We plot the T-vorticity \eqref{e:ext1-sp-sol-chi}, the corresponding temperature
$\theta(x)$ and the response vorticity \eqref{e:ext1-sp-sol-omega} in \figref{f:ext1-sp-sols}.
T-vorticity \eqref{e:ext1-sp-sol-chi} has a double-bump
for any non-vanishing $\chi_0'$. For this shape of the response
vorticity \eqref{e:ext1-sp-sol-omega}, the vorticity-flux balance
argument given in the previous section can be also applied.
In addition, we can obtain a localized $\tomega$ solution by adding the
homogeneous solution of Eq.\eqref{e:ext1-eq-res-omega} in a similar way
to the Burgers layer type solutions.

\begin{figure}
\iffigs
\includegraphics[width = 8.5cm]{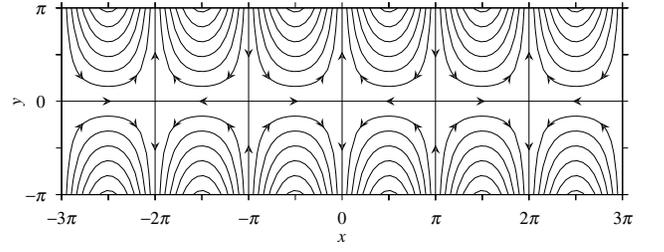} 
\else\drawing 65 10 {ext1_sin_field_arrow.eps}
\fi
\caption{\label{f:ext1_sin_field} Stream lines of an external flow $\mPsi(x) = -y\sin x$. The
 arrows denote the directions of the flow. Stagnation points are $(n \pi, 0), n = 0, \pm 1, \pm2, \cdots$.}
\end{figure}

As an another example, let us consider the periodic flow: 
\begin{equation}
f(x) = \sin x.
\label{e:ext1_sin} 
\end{equation}
This flow has vorticity $\mOmega = -Ay\sin x$. The corresponding forcing 
$F(x,y)$ is then $\nu A y\sin x$.
The streamlines of this external flow Eq.\eqref{e:ext1_sin} are
shown in \figref{f:ext1_sin_field}, where  stagnation points are 
located periodically.

\begin{figure}
\iffigs
\includegraphics[width = 8.5cm]{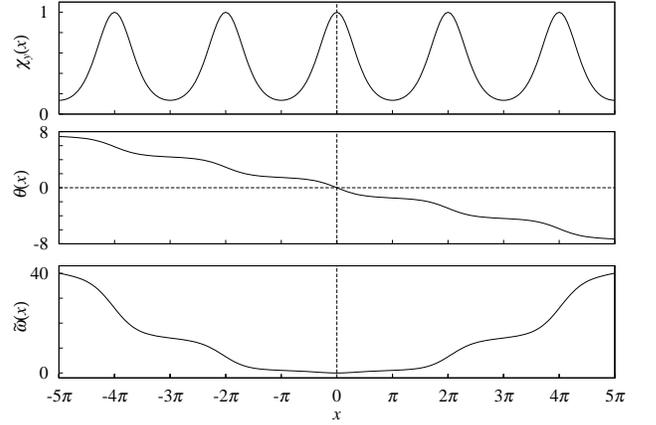} 
\else\drawing 65 10 {ext1_sin.eps}
\fi
\caption{\label{f:ext1_sin}FT type solution for $\mPsi(x) = - Ay\sin x$:
 the T-vorticity  Eq.\eqref{e:ext1_sin_sol_chi}, the corresponding temperature $\theta(x)$
 and the response vorticity  Eq.\eqref{e:ext1_sin_sol_tomega}. 
 Here we set the parameters as  $A = 1, \kappa = 1, \alpha g = 1, \varphi = \pi /4,
  \chi_0 = 1, \theta(x = 0) = 0$}
\end{figure}

For FT type solutions,  we impose the boundary conditions
\eqref{e:bc_ext_ftd_chi}-\eqref{e:bc_ext_ftd_dtomega}.
Then the general solutions \eqref{e:ext1-sol-chi} and
\eqref{e:ext1-sol-tomega} is expressed as
\begin{eqnarray}
 \chiy(x) &=& \chi_0 e^{-{A \over \kappa} ( 1 - \cos x)},
 \label{e:ext1_sin_sol_chi}\\
 \tomega(x) &=& \chi_0 \frac{\alpha g \sin \varphi}{A}
  \int_0^{x}  \xi  e^{-{A \over \kappa} (1 - \cos \xi)} d\xi.
 \label{e:ext1_sin_sol_tomega}
\end{eqnarray}
They are plotted in \figref{f:ext1_sin}. Multiple bumps in $\chiy(x)$ 
are seen. From the monotonic behavior of the temperature which
is similar to the Burgers T-vortex layer solution, these solutions
\eqref{e:ext1_sin_sol_chi} and \eqref{e:ext1_sin_sol_tomega} are
classified to FT solutions.
The remarkable feature of this solution is the staircase-like behavior of $\theta(x)$.
This kind of staircase-like change of the temperature is sometimes
observed in DNS.
Hence the solution \eqref{e:ext1_sin_sol_chi} and
\eqref{e:ext1_sin_sol_tomega} can be useful to describe the small
wavy behavior embedded on a temperature front.

\begin{figure}
\iffigs
\includegraphics[width = 8.5cm]{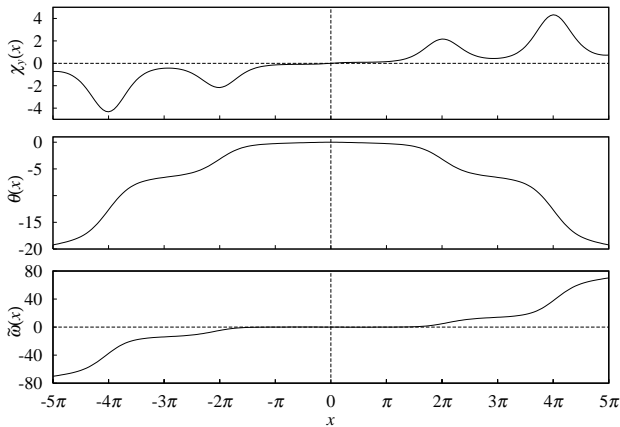} 
\else\drawing 65 10 {ext1_sin2.eps}
\fi
\caption{\label{f:ext1_sin_itd}IT type solutions for $\mPsi(x) = - Ay\sin x$;
 the T-vorticity Eq.\eqref{e:ext1_sin_itd_chi}, the corresponding
 temperature $\theta(x)$ and the vorticity Eq.\eqref{e:ext1_sin_itd_tomega}. 
 Here we set the parameters as  $A = 1, \kappa = 1, \alpha g = 1, \varphi = \pi /4,
  \chi_0'=  0.1$ and $\theta(x = 0) = 0$.}
\end{figure}

To obtain IT type solutions, we use the boundary conditions
\eqref{e:bc_ext_itd_chi}-\eqref{e:bc_ext_itd_dtomega}.
The solutions \eqref{e:ext1-sol-chi} and \eqref{e:ext1-sol-tomega} are
then expressed as
\begin{eqnarray}
 \chiy(x) &=& \chi_0' e^{-{A \over \kappa} ( 1 - \cos x)}
  \int_0^x e^{{A \over \kappa} ( 1 - \cos \xi)} d\xi,
 \label{e:ext1_sin_itd_chi}\\
 \tomega(x) &=& \chi_0' \frac{\alpha g \sin \varphi}{A}
  \int_0^{x}  d\zeta   e^{-{A \over \kappa} (1 - \cos \zeta)} \nonumber \\
 &&\times  \int_0^{\zeta} d\mu \int_0^{\xi} d\xi
   e^{{A \over \kappa} (1 - \cos \xi)}.
 \label{e:ext1_sin_itd_tomega}
\end{eqnarray}
These IT type solutions are shown in \figref{f:ext1_sin_itd}.
Its temperature behavior is similar to the Burgers IT solution.
The vorticity-flux balance argument can explain the shape of $\tomega$
in this case as well. However the response vorticity
\eqref{e:ext1_sin_itd_tomega} does not seem to
converge as $x \to \pm\infty$, it would be no use considering the
localized response vorticity in this case.

\section{concluding remarks}
\label{s:cr}
We have examined a class of steady solutions to the two-dimensional
Boussinesq approximation equations. We believe that these solutions well
describe coherent structures observed in DNS of 2DFC.
Under an extended stagnation flow $\Psi=-Ayf(x)$, 
we obtained two types of exact solutions; FT and IT types.
In both cases, T-vortex layers are maintained steady by the balance between 
the strain of the external stagnation flow and the diffusion of
T-vorticity. Furthermore,  in the FT case  a pair of T-vortex layers with
opposite signs are annihilated by the compression of the extended
stagnation flow without the continuous supply of T-vorticity which is 
maintained by the infinity of  temperature at $x=\pm \infty$.

For the former (FT) type including the Burgers vortex layer type 
solution when $f(x)=x$,
temperature remains  finite as $x$ goes to $\pm \infty$ and both
T-vorticity and vorticity are localized around the origin. 
In contrast, for the latter (IT) type, the absolute value of temperature 
should diverge as $x$ goes to $\pm \infty$ as mentioned above; 
T-vorticity tends to zero quite slowly and  vorticity is saturated to 
a finite value.  If $f(x)=x$, T-vorticity takes at most two bumps. 
On the other hand, if $f(x)\ne x$ then  it  may take as many bumps as possible.
although  we only dealt with  $f(x)=4x^3 -2x$ 
 and $f(x)=\sin\varphi$ where two and infinite number of bumps are 
observed, respectively. These bumps seem to correspond to fine T-vortex
 structures observed in DNS of 2DFC turbulence.

So far we have taken for granted the independence of an external flow 
and a response. This view may be reasonable if we focus on local
behavior  of the coherent structures. Indeed the local shape of 
the structures are quite similar to the steady solutions.
Thus it is  suggested that the characteristic time scale of the
external strain field are well separated  from that of structure's motions.
The role of the coherent structures  on  statistical properties of
turbulence  expected to be universal, is  still an open question. 
Since the coherent structures  have quite long correlation length 
comparable to the scale of the energy (temperature variance,
i.e. entropy)  containing range, their existence may break the locality 
 hypothesis significantly. However,  the authors have examined 
 the relative diffusion in 2DFC turbulence recently and found 
that the coherent structures rather play an essential role in
keeping the locality. These contradicting characteristics of the 
coherent structures make research on turbulence challenging.

In this sense, free convection turbulence is quite similar to
3DNS turbulence. We believe that coherent structures are universal 
and essential ingredients of turbulence. Therefore extensive researches 
on coherent structures are required.
 
\begin{acknowledgements}
This work has been partially supported by Grant-in-Aid for Science Research 
on Priority Areas (B) from the Ministry of Education, Culture, Sports, 
Science and Technology of Japan.
Numerical computation in this work was carried out at the 
Yukawa Institute Computer Facility. 
\end{acknowledgements}

\end{document}